\documentclass[journal,draftcls,onecolumn,11pt]{article}
  \usepackage{epsfig}
\begin{document}
\title{Maximum lifetime broadcasting problem in sensor networks}
\author{
Z. Lipi{\'n}ski \\
   Institute of Mathematics and Informatics \\
   Opole University,  Poland \\
}
\maketitle
\begin{abstract}
We solve the maximum lifetime problem for the point-to-point and point-to-multipoint
broadcast data transmission in one dimensional regular sensor network.
Based on the analytical solution of the problem for one dimension
we propose an algorithm solving the maximum lifetime broadcasting problem
for point-to-point data transmission for any dimension.
\end{abstract}
\section{Introduction}
A sensor network is type of a wireless ad-hoc network which nodes have limited power and
computational resources.
Typical activity of sensor network nodes is collection of sensed data,
performing simple computational tasks and transmission of the resulting data to a fixed set of data collectors.
These activities are energy consuming and require cooperation between nodes so they can share their energy
and other resources.
The sensors utilize most of their energy in the process of data transmission,
this energy grows with the size of the network and the amount of data transmitted over the network.
One of the most important problems in sensor networks is to
optimize the energy consumed by the sensor network nodes
to extend their operating time and extend the lifetime of the whole network.
By a sensor network lifetime we mean the time until first node of the network runs out of energy, \cite{Giridhar, Acharya}.
If the batteries of the nodes have initial energy $E_0$, then by finding the optimal energy utilization
by each node $E_i^{{\rm opt}}$ we can determine the number of cycles
$N_{{\rm cycles}}=[\frac{E_0}{E_{i'}^{{\rm opt}}}]$
the network can perform its functions until the most overloaded node runs out of its energy $E_{i'}^{{\rm opt}}$.

In wireless networks one can distinguish two types of data transmissions:
point-to-point and point-to-multipoint.
In the point-to-point transmission the sender transmits data to a unique receiver.
In the point-to-multipoint transmission the transmitter sends in parallel the same data to a set of receivers.
Important future of the point-to-multipoint data transmission is the wireless multicast advantage property
(the WMA property) \cite{Wieselthier}.
For such transmission the nodes, which are in the range of the transmitting node, can receive the data
without additional cost for the transmitter.
In this paper we analyze the maximum lifetime broadcasting problem for wireless ad-hoc networks
in which nodes use the point-to-point and point-to-multipoint data transmission methods.
We analyze the optimal broadcast data transmission in the one dimensional, regular sensor network $L_N$,
for which the data transmission cost energy matrix $E(x_i,x_j)$ is an arbitrary super-additive function, i.e.,
it satisfies the inequality
\begin{equation} \label{super-additive}
\forall_{x_i \leq x_j\leq x_k}\;E(x_i,x_j) + E(x_j,x_k)\leq E(x_i,x_k),
\end{equation}
where $x_i$ is the location of the network nodes on the line $R$.
The matrix $E(x_i,x_j)$ defines the transmission cost of one unit of data between $i$-th and $j$-th node.
Based on the analytical solution of the maximum lifetime broadcasting problem for $L_N$ network,
we propose a point-to-point broadcast data transmission algorithm which allows to maximize the lifetime of
sensor networks in two and more dimensions.

Let us denote by $S_N$ a one dimensional sensor network build of $N$ nodes.
The nodes are located at the points $x_i$, $i\in [1,N]$ of a real line.
By $Q_k$ we denote the amount of data which generates and broadcasts the $k$-th node.
The data transmission flow matrix $q^k(x_i,x_j) \equiv q^k_{i,j}$ defines the amount of data which is sent
by the $i$-th node to the $j$-th node in the process of data broadcasting.
If the $k$-th node broadcasts the amount $Q_{k}$ of data
then the energy spent by the $i$-th node to retransmit this data is given by the formula
\begin{equation} \label{NodeEnergy}
E_{i}(q^{k}) = \sum_{j=1, j\neq i}^{N} q_{i,j}^{k} E_{i,j}.
\end{equation}
The requirement, that each node must receive $Q_k$ of data can be written in the form
\begin{equation} \label{DataFlowConstrain}
 \forall_{j\in [1,N], j\neq k} \;\sum_{i=1, i\neq j}^{N} q_{i,j}^{k} = Q_k.
\end{equation}
To solve the maximum lifetime broadcasting problem means to find
a data transmission flow matrix $q_{i,j}^{k}$, such that the objective function
\begin{equation} \label{ObjectiveFunction}
E_{i'}(q^{k}) = \max_{i} \{ E_{i}(q^{k}) \},
\end{equation}
reaches its minimum. 
The maximum lifetime broadcasting problem, later referred to as MLB problem, is defined by the set of
parameters: the number of nodes $|S_N|=N$, the amount of data $Q_k$ which is broadcasted by the $k$-th node, $k\in [1,N]$
and the data transmission cost energy matrix $E_{i,j}$.
If the matrix $E_{i,j}$ is a function of a distance $d(x_i,x_j)$ between two nodes,
then instead of defining the matrix elements $E_{i,j}$
one can give the location of network nodes
in the space and determine the values of $E_{i,j}$ from the metric $d(x_i,x_j)$, $i,j\in [1,N]$.
In this paper, we assume that $E_{i,j}$ is an arbitrary polynomial function of the distance $d(x_i,x_j)$
and satisfies (\ref{super-additive}).

We represent a sensor network $S_N$ as a directed, weighted graph $G_N=\{S_N, V, E \}$ in which
$S_N$ is a set of graph nodes, $V$ is the set edges and $E$ set of weights.
Each directed edge $t_{i,j}\in V$ defines a communication link between $i$-th and $j$-th node of the network.
To each edge $t_{i,j}$ we assign a weight $E_{i,j}$, which is the cost of transmission of one unit of data between
$i$-th and $j$-th node.
By $U^{({\rm out})}_{i} \subseteq S_N$ we denote a set of the network nodes
to which the $i$-th node can sent the data
$$U^{({\rm out})}_{i}=\{ x_j\in S_N | \;\exists \;t_{i,j}\in V \}.$$
The set $U^{({\rm out})}_{i}$ defines the maximal transmission range of the i-th node.
In the paper we assume, that each node of $S_N$ can send data to any other nodes of the network, i.e.,
\begin{equation} \label{TheSetUFullRange}
\forall_{i\in [1,N]}, \;\; U^{({\rm out})}_{i}=S_N.
\end{equation}
The data flow in the network $S_N$ we describe in terms of spanning trees of the graph $G_N$.
The set of all spanning trees, which begin in the $k$-th node we denote by $V_{k}^{}$.
If the assumption (\ref{TheSetUFullRange}) is satisfied,
then $G_N$ is a complete graph and the number of spanning trees rooted at k-th node $|V_{k}^{}|$ it is equal to $N^{N-2}$, \cite{Berge}.
We will identify a tree $t^{k,r}$ with the set of its edges.
Because each tree $t^{k,r}$ defines the data flow from the broadcasting node to all other nodes of the network then they
must have a direction and must be ordered in a sequence that the data is transmitted along the tree.
We assume, that edges of the tree $t^{k,r}\in V_k$ have the natural orientation from the root node to the root child, etc., and are arranged with some fixed, initial sequence
\begin{equation} \label{TreeInitialSequence}
t^{k,r}=(t^{k,r}_{i_1,j_1},..., t^{k,r}_{i_{N-1},j_{N-1}}).
\end{equation}
For a point-to-point data transmission the solution of maximum lifetime broadcasting problem
does not depend on the ordering of the edges.
The ordering is important for the point-to-multipoint broadcast data transmission,
where different wireless multicast data transmissions are defined
by different orderings of the edges in the tree $t^{k,r}_{}$.

By $q^{k,r}_{i,j}$ we denote the amount of data which is transmitted along the edge $t^{k,r}_{i,j}$.
The data $q_{i,j}^{k}$ which is sent by the $i$-th node to the $j$-th node along all trees is given by the formula
\begin{equation} \label{qijtkr}
q_{i,j}^{k} = \sum_{r=1}^{N^{N-2}} q^{k,r}_{i,j} t^{k,r}_{i,j}.
\end{equation}
We require, that along each tree the transmitted data is the same,
which means that for fixed $k$ and $r$ the weights of the edges $t^{k,r}_{i,j}$ are equal,
$\forall_{i,j,i',j'}\;q^{k,r}_{i,j} = q^{k,r}_{i',j'}$.
We denote these weights by $q^{k}_{r}$, i.e.,
\begin{equation} \label{t-q}
\forall_{i,j} \; q^{k,r}_{i,j} = q^{k}_{r}.
\end{equation}
The objective function of the MLB problem (\ref{ObjectiveFunction})
can be written in terms of trees $t^{k,r}$ and weights $q^{k}_{r}$ with the help
of the following formula for the energy of each sensor
\begin{equation} \label{NodeEnergy1}
E_{i}^{k} = \sum_{j=1, j\neq i,k}^{N} \sum_{r=1}^{N^{N-2}}  q^{k}_{r} \; t^{k,r}_{i,j}\; E_{i,j}.
\end{equation}
The requirement (\ref{DataFlowConstrain}) written in terms of  $t^{k,r}$ and $q^{k}_{r}$ has the form
\begin{equation} \label{sumqk=Qk}
\sum_{r} q^{k}_{r} = Q_k.
\end{equation}
With the help of (\ref{NodeEnergy1}) and (\ref{sumqk=Qk}) the maximum lifetime broadcasting problem
can be defined by the set of the following formulas
\begin{equation} \label{DefinitionOfMLBTPForTrees}
\left\{   \begin{array}{l}
\min_{q^{k}} \max_{i\in L_N} \{ E_{i}^{k}(q^{k}) \}_{i=1}^{N}, \;k\in [1,N], \\
E_{i}^{k}(q^{k}) = \sum_{j=1}^{N} \sum_{r}^{} q^{k}_{r}\; t^{k,r}_{i,j}\; E^{}_{i,j},\;\; E^{}_{i,j}\geq 0,\\
t^{k,r} \in V_{k}^{}, \;q^{k}_{r}\geq 0,\; r\in [1,N^{N-2}], \\
\sum_{r} q^{k}_{r} = Q_k,\\
\end{array} \right.
\end{equation}
where $q^{k}=(q^{k}_{1}, ..., q^{k}_{N^{N-2}})$.
A solution of the MLB problem (\ref{DefinitionOfMLBTPForTrees}),
when the data is broadcasted by the $k$-th node,
will be given by an optimal set of trees  $\{ t^{k,r}_{} \}_{r=1}^{N'} \subset V_{k}^{}$, $N'\leq N$,
weighted by $N'$ non-negative numbers $q^{k,{\rm opt}}_{r}$,
which satisfy (\ref{sumqk=Qk}) and minimize the objective function (\ref{ObjectiveFunction}).
The following lemma describes relation between the local minima of the objective function (\ref{ObjectiveFunction})
and global solutions of (\ref{DefinitionOfMLBTPForTrees}).

{\bf Lemma 1.}
 Any minimum of the objective function (\ref{ObjectiveFunction}) is a solution of the maximum lifetime broadcasting problem.

{\it Proof.}
The feasible set
$$ U(q^k)=\{ q^k \in R_{+}^{N^{N-2}}| \sum_{r=1}^{N^{N-2}} q^{k}_{r} - Q_k=0 \},$$
where $R_{+}$ is a set of non-negative real numbers and
is a domain of the objective function $E(q^{k})$.
From the linearity of the set $U(q^k)$ it follows, that any two points $q^k_0, q^k_1\in U_(q^k)$ can be connected by a line segment and
$\forall_{t\in [0,1]}\; tq^k_0 +(1-t)q^k_1 \in U(q^k).$
The objective function (\ref{ObjectiveFunction}) of the maximum lifetime broadcasting problem (\ref{DefinitionOfMLBTPForTrees})
is piecewise linear in the $q^k$ variable, which means that
$E(tq^k_0 +(1-t)q^k_1)= t E(q^k_0) +(1-t) E(q^k_1).$
Let us assume, that $q^k_0$ is a minimum of $ E(q^k)$ and there exists another local minimum $q^k_1$
such that
$E(q^k_0)< E(q^k_1),$
then from the linearity of $ E(q^k)$ it follows that
$\forall_{t\in [0,1]}\;\; E(t q^k_0 +(1-t)q^k_1)= t E(q^k_0) +(1-t)E(q^k_1) < tE(q^k_1) +(1-t)E(q^k_1)$
and $\forall_{t\in [0,1]}\;q^k(t)$, the inequality
$E(q^k(t))< E(q^k_1)$
holds, which  means that $q^k_1$ cannot be a minimum of $E(q^k)$ in $U(q^k)$
and $q^k_0$ is a global minimum. $\diamond$

If we assume that $Q_{k}, \; q_{i,j} \in Z_{+}^{}$ in (\ref{DefinitionOfMLBTPForTrees}),
then we get the mixed integer linear programming problem for the network lifetime.
The following lemma describes the complexity of the problem.

{\bf Lemma 2}.
The maximum lifetime broadcasting problem is NP-hard.

{\it Proof.}
We reduce the partition problem to the MLB problem (\ref{DefinitionOfMLBTPForTrees}).
To solve the MLB problem we must find a set of weights $\{q^{k}_{r} \}_{r \in [1,N^{N-2}]}$
for all trees from $V_{k}$, which minimize the objective function (\ref{ObjectiveFunction}).
For the integer values of $q^{k}_{r}\in Z_{0}^{+}$, and $Q_k\in Z^{+}$ to find
the weights $q^{k}_{r}$, $r\in [1,N^{N-2}]$ satisfying $\sum_{r} q^{k}_{r} = Q_k$
it is equivalent to solving the partition problem, \cite{Garey}.
$\diamond$
%
\section{Solution of the MLB problem for $L_N$}
In one dimension, the superadditivity property (\ref{super-additive}) of the data transmission cost energy matrix $E_{i,j}$ means that, the lowest cost of delivery of a given amount of data
from the $i$-th node located at $x_i$ to the $j$-th node located at $x_j$ is reached
for the next hop data transmission along the shortest path.
By a shortest path we mean a distance between two nodes, $d(x_i,x_j)=|x_i - x_j|$.
It can be easily shown that in one dimension
for $E_{i,j}=|x_i-x_j|^a$, $x_i,x_j \in R_+$
the inequality (\ref{super-additive}) is satisfied for any $a\geq 1$.
From the above it follows that any function of the form
%
$E_{i,j}(\bar{a},\bar{\lambda}) = \sum_{n=0}^{\infty} \lambda_{n} |x_i-x_j|^{a_n}$,
%
where $\lambda_n\geq 0$, $a_n \geq 1$ also satisfies (\ref{super-additive}).
Nodes of a regular network $L_N$ are placed at the points $x_i=i$ of a line
and the distance between neighboring nodes is equal to one, $d_{i,i\pm 1}=1$.
We can assume, without loosing the generality of the problem (\ref{DefinitionOfMLBTPForTrees}) that,
$E_{i,i\pm 1}(\bar{a},\bar{\lambda})=1$, which is equivalent to the requirement
that $\sum_{n=0}^{\infty} \lambda_{n} =1$.
By $E_{r}$ we denote the cost of transmission of one unit of data between nodes which distance
is $d_{i,i+r}=r$, $i,r \in [1,N]$.
The function $E_{r}$ defines weight for all edges $t_{i,i+r}$ of a given transmission tree $t$ in $V_k$ and
it has the form
\begin{equation} \label{E[r]}
E_{r}(\bar{a},\bar{\lambda}) = \sum_{n=0}^{\infty} \lambda_{n} r^{a_n}, \;\; r \in [1,N],
\end{equation}
where $\sum_{n=0}^{\infty} \lambda_{n} =1$, $\lambda_n\geq 0$ and $a_n \geq 1$.
We show that for one dimensional, regular sensor network $L_N$ with the
data transmission cost energy matrix of the form (\ref{E[r]}) there exists
an equal energy solution of the maximum lifetime broadcasting problem for nodes which are inside the network, i.e. $i\in [2,N-1]$,
and non equal energy solution for the border nodes, $i=1,N$ of $L_N$.
The next lemma describes the solution of the MLB problem for the first and last node of the $L_N$ network.

{\bf Lemma 3.}
The solution of the maximum lifetime broadcasting problem (\ref{DefinitionOfMLBTPForTrees}) with $E_{i,j}$ satisfying (\ref{super-additive}),
for the first and last node of the $L_N$ network
 is given by the transmission graphs
$(t_{1,2},..., t_{N-1,N})$
and
$(t_{N,N-1},..., t_{2,1})$  
with the weights $Q_1$ and $Q_N$ respectively.

{\it Proof.}
From the superadditivity property (\ref{super-additive}) of the data transmission cost energy matrix $E_{i,j}$ it follows that,
the nodes consume minimal energy when they send all of their data to the nearest node.
Because the distance between neighboring nodes of the regular network $L_N$ is the same,
then the energy spent by each node to transmit a given amount of data to its neighbor is also the same.
From the above it follows that, when the broadcasted data $Q_1$ is transmitted along the edges $t_{i,i+1}$, $i\in [1,N-1]$, then each node $i\in [1,N-1]$
spends the same, minimal amount of energy.
This means that the optimal transmission, a solution to the MLB problem for the first node is given by the data transmission graph $(t_{1,2} ,..., t_{i,i+1},..., t_{N-1,N})$ with the weight $Q_1$.
In a similar way one can find the solution of (\ref{DefinitionOfMLBTPForTrees}) for the last node of the $L_N$ network.
It is given by the transmission graph $(t_{N,N-1} ,..., t_{i,i-1},..., t_{2,1})$ weighted with the $Q_N$ of data. $\diamond$

The next theorem describes the solution of (\ref{DefinitionOfMLBTPForTrees}) for the one dimensional, regular
sensor network $L_N$, when the data is broadcasted by the internal nodes, $k\in [2,N-1]$ of the network.

{\bf Theorem  1.}
The solution of the maximum lifetime broadcasting problem (\ref{DefinitionOfMLBTPForTrees}), with the data cost energy matrix
of the form (\ref{E[r]}), when the data is broadcasted by the k-th node, $k\in [2,N-1]$,
is an equal energy solution and it is given by the $N$ transmission trees $t^{r}$, $r\in [1,N]$
\begin{equation} \label{OptimalTree2-N-Minus1-ver5}
   \begin{array}{ll}
 (t^{r}_{k,k-1},...,t^{r}_{2,1}, t^{r}_{r,k+1},t^{r}_{k+1,k+2},..., t^{r}_{N-1,N}),  &r\in [1,k-1],\\
 (t^{k}_{k,k-1},..., t^{k}_{2,1}, t^{k}_{k,k+1},..., t^{k}_{N-1,N}\},                  &r=k,\\
(t^{r}_{k,k+1}, ..., t^{r}_{N-1,N}, t^{r}_{r,k-1}, t^{r}_{k-1,k-2},..., t^{r}_{2,1}),  & r \in [k+1,N], \\
\end{array} \end{equation}
with the weights
\begin{equation} \label{q^k_i}  
q^k_r  = \left\{   \begin{array}{ll}
  \frac{E_1}{E_k} q^k_k + \frac{E_1}{E_k} Q_k,             & r=1, \\
  \frac{E_1}{E_{k+1-r}} q^k_k,                            & r\in [2, k-1], k\geq 3,\\
  \frac{E_1}{E_{r-k+1}} q^k_k,                            & r\in [k+1, N-1], k\leq N-2, \\
  \frac{E_1}{E_{N-k+1}} q^k_k + \frac{E_1}{E_{N-k+1}} Q_k, & r=N, \\
\end{array} \right.\end{equation}
where 
\begin{equation} \label{q^k_k}
q^k_k = \frac{1 - \frac{E_1}{E_k} - \frac{E_1}{E_{N-k+1}} }
             {-1+ \sum_{i=1}^{k} \frac{E_1}{E_i}+ \sum_{i=1}^{N-k+1} \frac{E_1}{E_{i}} } \;Q_k.
\end{equation}
{\it Proof.}
We assume that the formulas (\ref{OptimalTree2-N-Minus1-ver5}) and (\ref{q^k_i})
are solution to the MLB problem for $N$ and show that it is also true for $(N+1)$.
After adding the $(N+1)$-th node at the end of the network,
the nodes from the $L_{N}$ subnetwork can transmit data to the $(N+1)$-th node by a trees $t^{r'}(L_{N+1})$,
$r'\in [1, N']$ such, that  $t^{r'}(L_{N+1})|_{L_N}=t^{r}$ and $t^{r}$ are given by (\ref{OptimalTree2-N-Minus1-ver5}).
Because all $Q_{k}$ of data must be delivered to the $(N+1)$-th node by the nodes from the $L_N$ subnetwork,
then for data transmission cost energy matrix satisfying (\ref{super-additive})
the lower costs of delivery of these data
is when the $N$-th node passes all $Q_{k}$ of data
to the $(N+1)$-th node by the 'next hop' data transmission, consuming the amount
$$E^k_{N}=E_1 Q_{k}$$
of energy.
This means that the optimal trees $t^{r'}(L_{N+1})$ for data transmission from the $L_N$ subnetwork
are given by the ordered set of edges
\begin{equation} \label{TreeLN+1}
t^r(L_{N+1}) = ( t^{r},t^{r}_{N,N+1}), \;\;r\in [1,N], i,j\in [1,N],
\end{equation}
where the edges of the $t^{r}$ tree are given by (\ref{OptimalTree2-N-Minus1-ver5}).

Because, the $(N+1)$-th node must take part in data retransmissions, we must find set of trees which begin in the
$k$-th node and for which the $(N+1)$-th node is not a leaf.
From (\ref{TreeLN+1}) follows that all $Q_{k}$ of data is transmitted to the $(N+1)$-th node
in the $L_N$ subnetwork by the next hop data transmission graph, starting from the $(k+1)$-th node.
It means that the first $(N+1-k)$ edges of the optimal trees $t^{r''}$, $r''\in [N+1, N']$ are
\begin{equation} \label{N+1-NodeTree}
t^{r''}=( t^{r''}_{k,k+1}, ...,t^{r''}_{N,N+1}, ... ).
\end{equation}
The nearest node to which the $(N+1)$-th node can send the $q^{k}_{r''}$ data, without transmitting data in a loop,
is the $(k-1)$-th node, which means that next edge, which we must add to the set (\ref{N+1-NodeTree}) is
$t^{r''}_{N+1,k-1}$.
The next edges in the transmission trees (\ref{N+1-NodeTree}) are
$t^{r''}_{k-i,k-i+1}$, $i\in [1,k-1]$,
and define the most efficient
data transmission for $E_{i,j}$ satisfying (\ref{super-additive}).
From the above it follows, that the optimal data transmission tree for the $(N+1)$-th node to retransmit
part of the broadcasted data in the $L_{N+1}$ network is given by the formula
\begin{equation} \label{N+1-NodeTree2}
t^{N+1} = (t^{N+1}_{k,k+1}, ...,t^{N+1}_{N,N+1}, t^{N+1}_{N+1,k-1}, t^{N+1}_{k-1,k-2}, ..., t^{N+1}_{2,1} ),
\end{equation}
which is the $(N+1)$ tree in (\ref{OptimalTree2-N-Minus1-ver5}) for the $L_{N+1}$ network.

We show that it is enough to add one tree $t^{N+1}$ to the solution of MLB problem in $L_N$ network
to obtain the set of trees which is a solution for $L_{N+1}$.
For the set of trees $\{(t^{r}, t^{N+1})\}_{r=1}^{N}$  given by (\ref{OptimalTree2-N-Minus1-ver5}) and (\ref{N+1-NodeTree2})
and weights  $\{(q^{k}_{r},q^{k}_{N+1})\}_{r=1}^{N}$ an equal energy requirement can be written in the form
\begin{equation} \label{EqualEnergyEq}
E_{i}(q^{k}) = E_{i+1}(q^{k}), \;\; i\in [1,N],
 \end{equation}
where
$$ E_{i}(q^{k})  =  \left\{   \begin{array}{l}
q^k_{1} E_{k},\; i=1, \\
q^k_{i} E_{k+1-i} + Q_k E_1, i\in [2,k], \\
q^k_{i} E_{i-k+1} + Q_k E_1, i\in [k+1, N-1], k \leq  N-1, \\
q^k_{N} E_{N-k+1} + Q_k E_1,\; i=N, \\
q^k_{N+1} E_{N-k+2},\; i=N+1. \\
\end{array} \right.$$
It is easy to check that the solution of the equations (\ref{EqualEnergyEq}),
with the requirement $\sum_{r=1}^{N+1} q^{k}_{r} = Q_k$,
is given by (\ref{q^k_i}) for the $L_{N+1}$ network. $\diamond$

\begin{figure} [!ht] \begin{center}
\includegraphics[width=180pt]{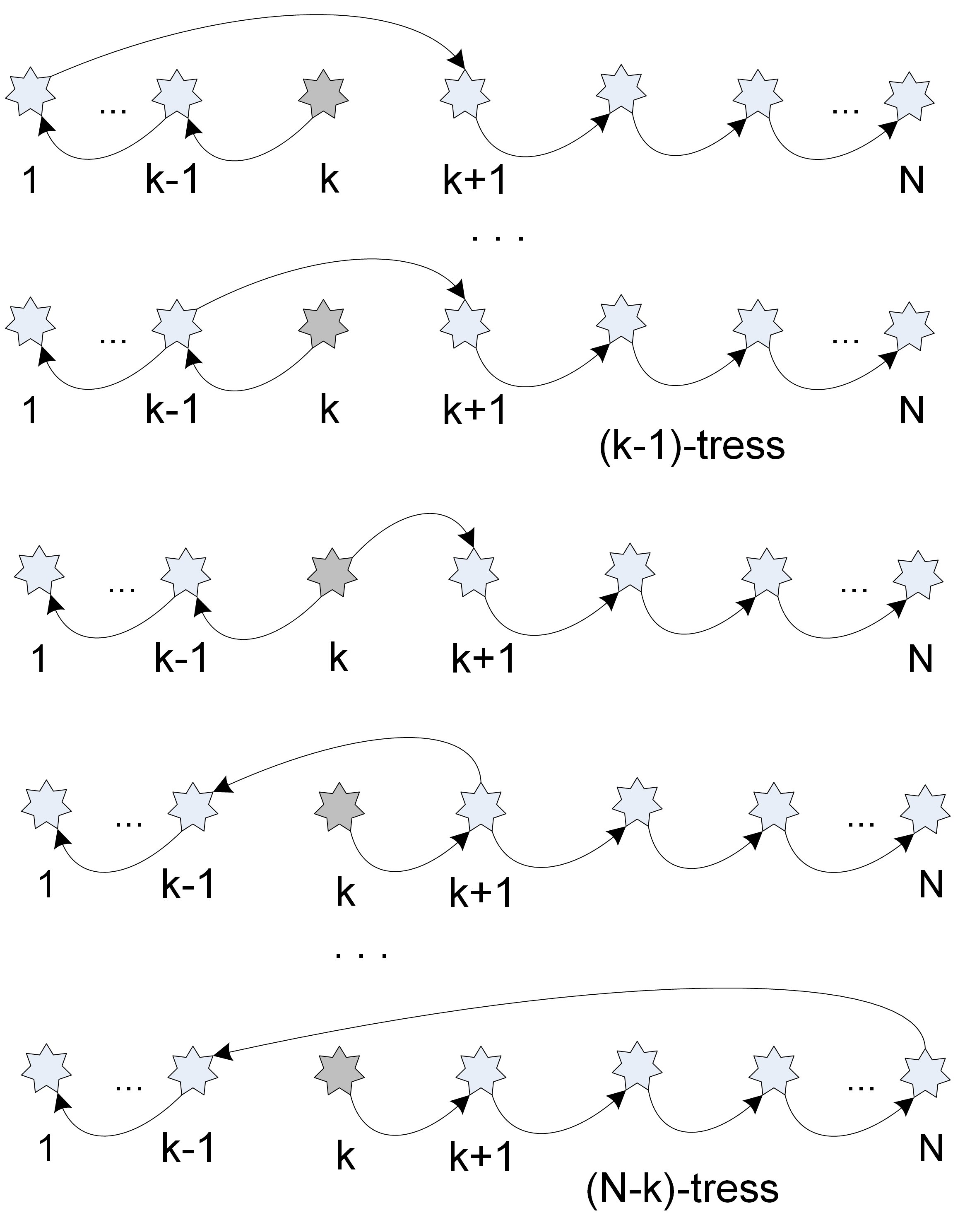}
\caption{The set of trees for a solution of the MLB problem in $L_N$.}
\label{SolutionTreesMLBP}
\end{center} \end{figure}

The Figure \ref{SolutionTreesMLBP} shows the optimal set of $N$ trees for
a solution of the MLB problem in $L_N$ when the data is broadcasted by the $k$-th node, $k\in [2,N-1]$.
%

The next lemma shows that for a sufficiently large network $L_N$, when the number of nodes grows,
the energy $E_i^{k}$, $k\in [2,N-1]$ of each node decreases.

{\bf Lemma 4}.
For the solution of the MLB problem given by (\ref{OptimalTree2-N-Minus1-ver5})-(\ref{q^k_k})
the energy of each node $E_{i}^{k}(L_{\infty})$, $i\geq 1$, $k \geq 2$ is given by
 \begin{equation} \label{EikLimitLN}
E_{i}^{k}(L_{\infty}) = (1+ \frac{1 - \frac{E_1}{E_k} }{ \sum_{r=1}^{k} \frac{E_1}{E_r} + \sum_{r=1}^{\infty} \frac{E_1}{E_{r}}}) E_1 Q_k.
\end{equation}
{\it Proof.}
The solution (\ref{OptimalTree2-N-Minus1-ver5})-(\ref{q^k_k})
is an equal energy solution with the energy of each node
$E^k_i(L_N) = E_1 q^k_k(L_N) + E_1 Q_k$, $i\in [1,N]$, $k\in [2,N-1]$,
where $q^k_k(L_N)$ is given by (\ref{q^k_i}).
For $N \rightarrow \infty$ the energy $E^k_i(L_N)$ has the limit (\ref{EikLimitLN}) $\diamond$.

\section{Broadcast transmission with wireless multicast advantage}
\begin{figure} [!ht] \begin{center}
 \includegraphics[width=200pt]{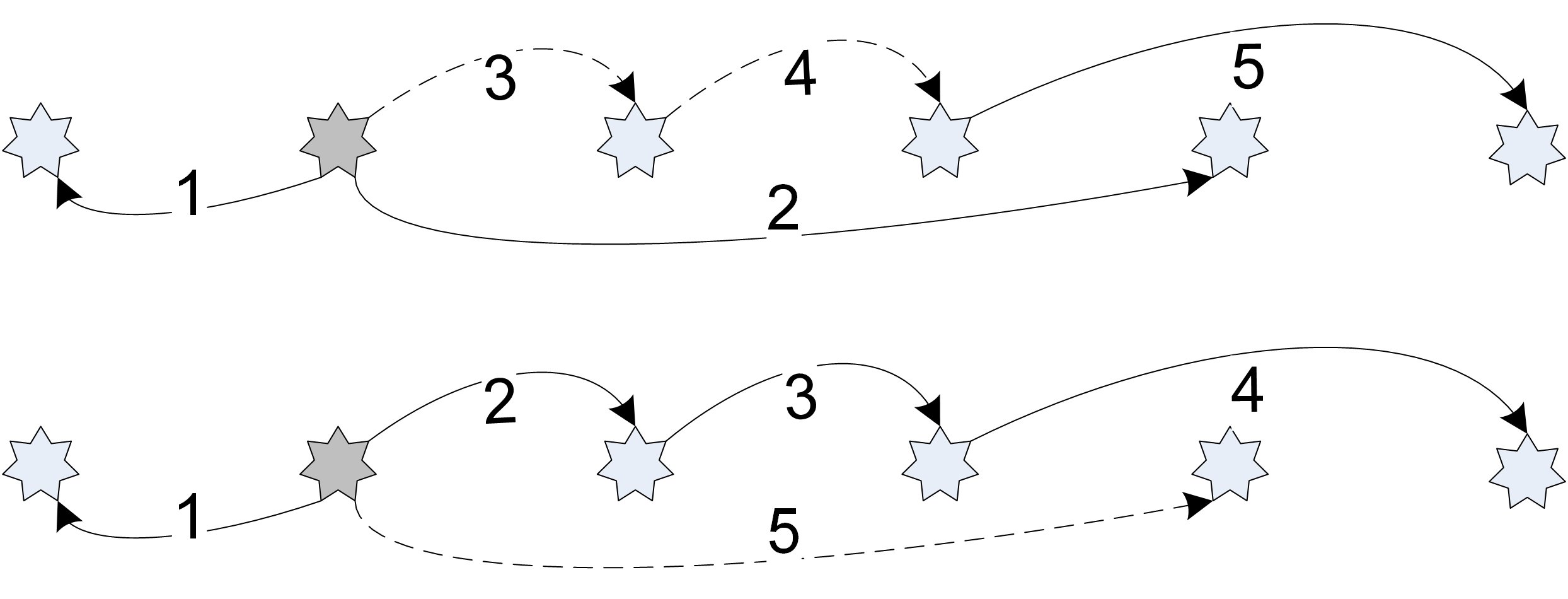}
\caption{Two nonequivalent WMA data transmissions along the same tree in $L_6$.}
\label{Fig-WMA-L6-TwoTrees}
\end{center} \end{figure}

Because any data broadcasted by the $k$-th node of the $L_N$ network is transmitted along
some spanning tree $t^{k,r}\in V_k$, we define the wireless multicast advantage property
of a broadcast data transmission for each tree of the set $V_k$.
In the first section we defined $V_k$ as the set of all spanning trees
of the graph $\{L_N, V, E \}$ rooted at the k-th node and we assumed that edges of each tree $t^{k,r}$
were arranged in some fixed sequence.
Because any WMA transmission along given tree $t^{k,r}$ is defined by some ordering of its edges $t^{k,r}_{i,j}$,
we extend the set $V_k$ by adding to it $(\alpha_r-1)$ trees $t^{k,r,\alpha}$ obtained by reordering
edges of the tree $t^{k,r}$.
Any ordered tree $t^{k,r,\alpha}$, $\alpha \in [1,\alpha_r]$,
where $t^{k,r,1}=t^{k,r}$ is an initial ordering (\ref{TreeInitialSequence}), defines some WMA data transmission.
The set of all trees $t^{k,r,\alpha}$, $\alpha \in [1,\alpha_r]$, $r\in [1,N^{N-2}]$
rooted at the k-th node we denote by $\overline{V}_k$.
In Figure \ref{Fig-WMA-L6-TwoTrees} there are two different orderings of a tree rooted at the second node
of the $L_6$ network shown, which define different WMA data transmissions.

The numbers on the edges denote the sequence the data is transmitted along the tree
and the doted lines indicate the costless transmission.

\begin{figure} [!ht] \begin{center}
 \includegraphics[width=200pt]{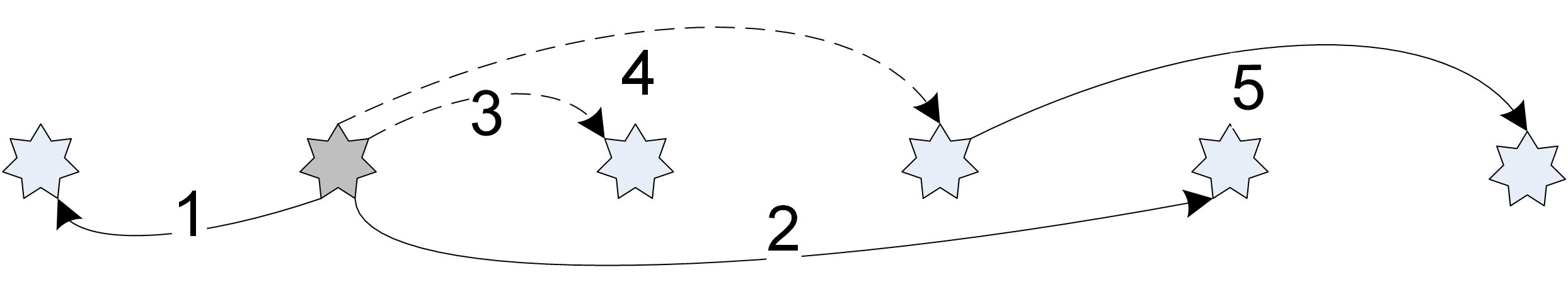}
\caption{Parallel WMA data transmission along a tree in $L_6$.}
\label{Fig-WMA-L6-firstTreeParallel}
\end{center} \end{figure}

Let us denote by $E_{i,j}(q)$ the energy consumed by the $i$-th node to transmit the amount $q$ of data
to the $j$-th node.
By $U^{}_{i,j}$ we denote  the set of network nodes which are in the transmission range of the $i$-th node,
when the data is transmitted to the $j$-th node, and can receive the transmitted data without additional cost of the transmitter
$$U^{}_{i,j} = \{ n \in S_N | \; E_{i,j}(q)>0 \Rightarrow  E^{}_{i,n}(q)=0 \}.$$
From the definition of $U^{}_{i,j}$ it follows, that the data transmitted
by the i-th node to the j-th node and nodes from $U^{}_{i,j}$ is parallel and
is performed along the edges $\{ t_{i,j}, t_{i,n}\}_{n \in U^{}_{i,j}}$ of some tree $t$.
Definition of a set $U(t^{k,r}_{i,j})\equiv U^{k,r}_{i,j}$ for each edge of a tree $t^{k,r} \in V_k$
is equivalent to the definition of some ordered tree $t^{k,r,\alpha}\in \overline{V}_k$, $\alpha \in [1,\alpha_r]$
and some WMA data transmissions along $t^{k,r}$.
It may happen that different trees $t^{k,r,\alpha}$ and $t^{k,r',\alpha'}$ from $\overline{V}_k$
define equivalent WMA data transmissions.
Figure \ref{Fig-WMA-L6-firstTreeParallel} showns an ordered tree in $L_6$
which defines WMA data transmission equivalent to the transmission along the first tree in Figure \ref{Fig-WMA-L6-TwoTrees}.

By $q^{k}_{r,\alpha}$ we denote the data transmitted along the ordered tree $t^{k,r, \alpha}\in \overline{V}_k$.
The WMA set $U^{}(t^{k,r, \alpha}_{i,j})\equiv U^{k,r,\alpha}_{i,j}$ for
the data transmission along the tree $t^{k,r, \alpha}$ we define as follows
$$U^{k,r,\alpha}_{i,j} = \{ n \in S_N| E^{}_{i,j}(q^{k}_{r,\alpha}) >0 \Rightarrow  \forall_{m\in S_N} E^{}_{m,n}(q^{k}_{r,\alpha})=0 \}.$$
We also assume that
 \begin{equation} \label{WMA-AddtionalAssumption}
 E^{}_{m,n}(q^{k}_{r,\alpha})=0 \Rightarrow U^{k,r,\alpha}_{m,n}=\emptyset,
 \end{equation}
which means that the WMA property cannot be applied to the data transmission for which $E_{m,n}(q)=0$
Note, that this assumption is non-trivial for the directional antennas.
The data transmission cost energy matrix for each edge of the tree $t^{k,r, \alpha}$,
$E^{}_{i,j}(t^{k,r, \alpha})\equiv E^{k,r,\alpha}_{i,j}$
having the WMA property and satisfying (\ref{WMA-AddtionalAssumption}) can be defined by the following formulas
\begin{equation} \label{WmaCostMatrixE}
E^{k,r,\alpha}_{i_n,j_n} =
\left\{   \begin{array}{l}
    E^{}_{i_1,j_1}, n=1,\\
    E^{}_{i_n,j_n}, \;  j_n \notin \bigcup_{s=1}^{n-1} U^{k,r,\alpha}_{i_s,j_s}, \;n\in [2,N-1],\\
    0, \;\; \;\; \;\; \;\; j_n \in \bigcup_{s=1}^{n-1} U^{k,r,\alpha}_{i_s,j_s}.\\
 \end{array} \right.
 \end{equation}
The energy consumed by the i-th node, written in terms of $E^{k,r,\alpha}_{i,j}$, equals
\begin{equation} \label{NodeEnergyWithWMA}
E_{i}(q^{k}) = \sum_{j=1, j\neq i,k}^{N} \sum_{r,\alpha}^{}  q^{k,\alpha}_{r} \; t^{k,r,\alpha}_{i,j}\; E^{r,\alpha}_{i,j}.
\end{equation}
The objective function of the MLB problem with WMA property has the form (\ref{ObjectiveFunction})
where $E_{i}(q^{k})$ is given by (\ref{NodeEnergyWithWMA}).
The following lemma describes a solution of the MLB problem with WMA property when
nodes of the $L_N$ network use the bidirectional antennas.

{\bf Lemma 5.}
For the bidirectional antenna the solution of the MLB  problem with the wireless multicast advantage property
and $E_{i,j}$ satisfying (\ref{super-additive}) is given by the transmission tree
\begin{equation} \label{SolutionForBidirectional}
(t_{k,k-1}^{k},...,t_{2,1}^{k}, t_{k,k+1}^{k},...,  t_{N-1,N}^{k})
\end{equation}
with the weights $Q_k$, $k\in [2,N-1]$.

 {\it Proof.}
Since the distance between neighboring nodes in the $L_N$ network is equal to one,
then for a bidirectional antenna the data transmitted by the $k$-th node, $k\in [2,N-1]$ to the $(k-1)$-th node
is delivered to the $(k+1)$-th node without costs.
In other words, for the WMA data transmission along the edges $(t_{k,k-1}^{k}, t_{k,k+1}^{k})$
the amount $Q_k$ of data is delivered to the both nodes $(k\pm 1) \in L_N$
with the minimal energy $E_{k}^{k}=Q_k$ consumed by the $k$-th node.
From Lemma 3 we know that, the optimal tree to transmit the amount $Q_k$ of data from the $(k-1)$-th node to
the nodes $i\in [1,k-2]$ of the $L_N$ network is given by the sequence of edges
$(t_{k-1,k-2}^{k},...,t_{2,1}^{k})$ and the energy utilized by each node is qual $E_{i}^{k}=Q_k$, $i\in [2,k-2]$ and $E_{1}^{k}=0$.
Similarly, the optimal tree to transmit the amount $Q_k$ of data from the $(k+1)$-th node to
the nodes $i\in [k+2,N]$ of the $L_N$ network is given by the sequence of edges
$(t_{k+1,k+2}^{k},...,  t_{N-1,N}^{k})$ and the energy utilized by each node is qual $E_{i}^{k}=Q_k$, $i\in [k+2,N-1]$ and $E_{N}^{k}=0$.
This optimal transmission tree coincides up to an ordering with the transmission tree given by (\ref{SolutionForBidirectional}).  $\diamond$

The next theorem states that for the directional antennas and
for the internal nodes of the network $L_N$,
i.e. $k\in [2,N-1]$, the solution of the MLB problem with WMA property
coincides with the solution given in (\ref{OptimalTree2-N-Minus1-ver5})-(\ref{q^k_k}).

{\bf Theorem 2.}
For a directional antenna the solution of the MLB problem with the wireless multicast advantage property
and $E_{i,j}$ satisfying (\ref{super-additive}), for $k\in [2,N-1]$, $N\geq 3$
is an equal energy solution
and it is given by (\ref{OptimalTree2-N-Minus1-ver5})-(\ref{q^k_k}).

{\it Proof.}
The proof is basically the as the proof of Theorem 1. $\diamond$

The solutions of MLB problem with the wireless multicast advantage property
for the first and the last node of the $L_N$ network and for both types of antennas,
the bidirectional and  directional is given by Lemma 3.

\section{The maximum lifetime broadcasting algorithm}
The objective of the algorithm is to find an optimal weighted graph
for the point-to-point broadcast data transmission which minimizes the energy consumed by the most overloaded sensor
and thus maximizes the lifetime of the network $S_N$.
In this section we assume, that the network $S_N$ is build of $N$ nodes and
it is embedded in $d\geq 1$ dimensional space $R^d$.
The algorithm consists of two parts.
First, we determine the optimal transmission graph $g^{k}$
for the data broadcasted by the $k$-th node in $S_N$.
In the second step, we minimize the objective function of the MLB problem (\ref{ObjectiveFunction})
with respect to the weights $\{ q^{k}_{r}\}$ of the graph  $g^{k}$ and
the constraint $\sum_{r}^{} q^{k}_{r} = Q_k$.
We assume, that the optimal graph $g^{k}$ is a set of $N$ subgraphs $g^{k,r}$, $r\in [1,N]$ and
each subgraph is a sum of a shortest path $p^{k,r}$ from the broadcasting node to the $r$-th node
and the minimum spanning tree $t^{k,r}$ in the graph $G_N \setminus \bar{p}^{k,r}$ rooted at the $r$-th node,
where $\bar{p}^{k,r}= p^{k,r} \setminus \{ x_r\}$, which the out-order of each node does not exceed two.
The assumption, that the maximum out-order of the nodes of $g^{k}$ is equal to two
follows from the observation that, copying the data unnecessary overloads the sensors,
and as a general rule we allow the nodes to send the same data only to the maximum of two nodes.
For each $g^{k,r}$, $r\in [1,N]$ we assign a weight $q^{k}_{r}$ which
defines the amount of data transmitted along this subgraph.
%
The objective function of the algorithmic MLB problem is given by (\ref{ObjectiveFunction}), where
the energy of each node
$ E_{i}(q^k)=\sum_{r, j}^{} q^{k}_r g^{k,r}_{i,j}$ is
expressed in terms of the optimal graph $g^{k}=\{ p^{k,r}\cup t^{k,r}\}_r$ and its weights $q^{k}_r$.

The maximum lifetime broadcasting algorithm for the point-to-point data transmission
is defined as follows:

{\bf Step 1.}
In the graph $G_N=\{S_N, V, E \}$ determine the set $V_{k,k}$ of ordered,
minimal spanning trees rooted at the $k$-th node with the maximum out-order of each node equal to two.
Select a one tree $t^{k,k}$ from the set $V_{k,k}$.
If the tree $t^{k,k}$ is a path,
then it is a solution of the algorithmic MLB problem with the weights $Q_k$,
i.e. $q^k_{i,j} = Q_k t^{k,k}_{i,j}$, else go to the Step 2.

{\bf Step 2.}
In $G_N$ determine the shortest path $p^{k,r}$
from the $k$-th node to each node $r\neq k$ of the network $S_N$.
For the subgraph $G_N \setminus \bar{p}^{k,r}$, where $\bar{p}^{k,r}= p^{k,r} \setminus  \{ x_r\}$,
determine the ordered, minimum spanning tree $t^{k,r}$, $r\in [1,N]$, $r\neq k$ rooted at the $r$-th node
with the maximum out-order of each node equal to two.
The graph, which is a sum of $p^{k,r}$ and $t^{k,r}$ we denote by $g^{k,r}$, $r\in [1,N]$. 

{\bf Step 3.}
To each graph $g^{k,r}$ assign a weight $q^{k}_{r}$
and find the minimum of the objective function $E(q^k)=\max_{i} \{E_{i}(q^k)\}$
with respect to the $N$ variables $q^{k}_{r}$ and with the constraint $\sum_{r=1}^{N} q^{k}_{r} = Q_k$,
where 
$E_{i}(q^k)=\sum_{r, j}^{} q^{k}_r g^{k,r}_{i,j}$.
For determined tuple of N numbers $q^k_{r}=(q^k_{1}, ...,q^k_{N})$, which minimizes
the objective function $E(q^k)$, the solution of the algorithmic MLB problem
can be given by the data flow matrix
$q^k_{i,j} = \sum_{r} q^k_r g^{k,r}_{i,j}$. $\diamond$

%
To determine the shortest path $p^{k,r}$ or the minimum spanning tree $t^{k,r}$ in a weighted graph $G_N$
we can apply a well known graph searching algorithms.
For example, the Dijkstra algorithm for finding the shortest path and
the Kruskal algorithm for finding the minimum spanning tree, \cite{Cormen}.
In general, for such algorithms elements of the weight matrix $E$ in $G_N=\{S_N, V, E \}$ can be arbitrary non-negative numbers $E(x_i,x_j)\geq 0$,
where $x_i$, $x_j$ are location of the sensors in $R^{d}$.
In the previous sections, we considered the MLB problem with the data transmission cost energy matrix $E(x_i,x_j)$,
which defines the weight matrix $E$ in $G_N$, to be a superadditive function.
In dimension $d\geq 2$
the weight function matrix $E_{r}(\bar{a},\bar{\lambda})$ given by (\ref{E[r]}) satisfy the superadditivity property
$E_{r+r'}\leq E_{r}+E_{r'}$ when $r,r'\in R^d_{+}$ and $\forall_{n}\; a_n\geq 2$.
For such weights the maximum lifetime broadcasting algorithm determines the same optimal
graph $g^{k,r}$ in $G_N$.
It means, that the only difference between solutions of the algorithmic MLB problem
for a given network $S_N$ and $E_{r}(\bar{a},\bar{\lambda})$ is given
by various weights $q^{k}_{r}$ determined in the third step of the algorithm.
The following lemma states that for the MLB problem in the $L_N$ network
the data transmission graph determined by the maximum lifetime broadcasting algorithm
is an exact solution of the MLB problem in $L_N$.

{\bf Lemma 6}.
The solution of the MLB problem (\ref{DefinitionOfMLBTPForTrees})
determined by the MLB algorithm
for the internal nodes of the $L_N$ network
is given by (\ref{OptimalTree2-N-Minus1-ver5})-(\ref{q^k_k}).

 {\it Proof.}
Because any data transmission cost energy matrix $E_{i,j}$ of the form (\ref{E[r]}) satisfies (\ref{super-additive})
then the minimum spanning tree determined by the algorithm rooted at the $k$-th node ($k\neq 1,N$) weighted by (\ref{E[r]})
coincides with the spanning tree $t^{k,k}$ given by the solution of the MLB problem in Theorem 1.
Each tree $t^{k,r}$, $r\neq k, r\in [1,N]$ from (\ref{OptimalTree2-N-Minus1-ver5})-(\ref{q^k_k})
is a sum of a shortest path $p^{k,r}$ and minimum spanning tree $\tilde{t}^{k,r}$
in the graph $L_N \setminus \bar{p}^{k,r}$.
For the weights of the graph $G_N=\{L_N, V, E \}$
given by the data transmission cost energy matrix $E_{i,j}$ of the form (\ref{E[r]})
the shortest path $p^{k,r}$ between the $k$-th node and $r$-th node of $L_N$ is given by the distance between them and
the minimum spanning tree in the subgraph $L_N \setminus \bar{p}^{k,r}$, where
$\bar{p}_{k,r}=p_{k,r}\setminus \{ r\}$ is given by (\ref{OptimalTree2-N-Minus1-ver5}) and
coincides with the spanning tree $\tilde{t}^{k,r}$ determined by the algorithm in $L_N \setminus \bar{p}^{k,r}$.
$\diamond$

When the data is broadcasted by the first and the last node of the $L_N$ network,
then the optimal transmission graph determined by the MLB algorithm
is a path and coincides with the solution of the MLB problem given in Lemma 3.
For a sensor network $S_N$ which forms a regular grid the weights $E_{i,j}$ are of the form (\ref{E[r]}),
the minimum spanning tree is the path and the algorithm gives and an exact solution of the MLB problem.
For several sensor network configurations in two dimensions the MLB algorithm gave for some nodes
of $S_N$ an exact solution and for other nodes the energy level was  $30\%$ worse with comparison to the exact solution, i.e.,
$\frac{E(q^{alg})}{E(q^{opt})} \simeq 1.3$.
\section{Conclusions}
In this paper we solved the maximum lifetime problem for a point-to-point and a point-to-multipoint
broadcast data transmission in one dimensional regular sensor network $L_N$.
Based on the analytical solution of the problem in $L_N$,
we proposed an efficient algorithm for finding a weighted graph
in a sensor network of any dimension and for the point-to-point broadcast data transmission
which allows to extend the network lifetime.
The main difference between the MLB algorithm and the known broadcasting algorithms is that
the proposed algorithm splits the broadcasted data into $N$ parts, sends each part along different transmission trees
and by this spreads more evenly the energy consumed by sensors of the network.
For evenly distributed sensors in a given area in $R^d$ one can expect that
there exits an equal energy solution of the maximum lifetime broadcasting problem.
For such networks the third step of MLB algorithm, in which we minimize the objective function
(\ref{ObjectiveFunction}), can be replaced by the problem of solving the following set of linear equations
$$ \left\{   \begin{array}{l}
\sum_{r}^{} q^{r}_{k} = Q_k, \\
\sum_{r,j}^{} q^{r}_{k} (g^{k,r}_{i,j} - g^{k,r}_{i+1,j}) = 0, \;i\in [1,N-1].
\end{array} \right.$$
Knowledge abut existence of an equal energy solution of the maximum lifetime broadcasting problem
greatly simplifies the solution of the problem.
It seems, that the most interesting, unsolved problem for the presented MLB algorithm
is to find a criteria for deciding whether for a given distribution of sensors in the space $R^d$
an equal energy solution of the maximum lifetime broadcasting problem exits.
%

\end{document}